\documentclass[review]{elsarticle}

\usepackage{lineno,hyperref}
\usepackage{longtable}
\modulolinenumbers[5]

\journal{Journal of \LaTeX\ Templates}

\begin{document}

\begin{frontmatter}

\title{Shaking during Ion-Atom Collisions}

\author[]{Prashant Sharma\corref{cor1}}
\ead{prashant@iuac.res.in}

\author[]{Tapan Nandi}

\ead{nandi@iuac.res.in}
\address{Inter University Accelerator Centre, Aruna Asaf Ali Marg, New Delhi - 110067, INDIA}

\cortext[cor1]{Corresponding author}


\begin{abstract}
Shaking (shakeup + shakeoff) probabilities accompanying ion-atom collisions are studied using hydrogenic wavefunctions for K-, L-, M- shell electrons in the sudden approximation limit. The role of recoil velocity in the shaking processes is discussed. Further, it is found that the suddenness of collision between projectile and target nuclei plays a major factor in shaking of respective atomic system than the recoil of nuclei. 
\end{abstract}

\begin{keyword}
\texttt{} Ion-atom collision \sep Electronic excitation \& ionization \sep Shaking process \sep Sudden approximation \sep Hydrogenic wavefunction

\end{keyword}

\end{frontmatter}

\linenumbers

\section{Introduction}
\label{sec1}
All nuclear and atomic events result in a change in the central potential and/or electronic environment of the corresponding atomic system. If the perturbation or net change in system occurs suddenly enough, then the orbital electrons may not respond so rapidly to rearrange themselves. This process subsequently leads to the electrons getting excited to an unoccupied bound state (shakeup process) or leaving the parent atom/ion (shakeoff process). This process is called shaking process \cite{Freedman1974}. Theoretically shaking processes are treated under the sudden approximation limit \cite{ff}. It is worth to note that a perturbation is called sudden if the time period $\tau$ of perturbation is less than that of periodic motion, $2\pi\omega_{n}^{-1}$ , of the orbital electrons \cite{Feoktistov2010}. There are many nuclear processes e.g. $\beta$-decay \citep{Levinger1953,Couratin2012}, $\alpha$-transfer reactions \citep{Sharma2015}, positron decay \citep{Cho1997}, internal conversion \citep{PhysRevC.11.1353,PhysRevC.3.2246} as well as atomic processes e.g. photoionization \citep{Mukoyamaa2010}, inner-shell ionization \citep{Mukoyama2005}, orbital electron capture \citep{Cho19972,PhysRevC.25.2722} etc. in which the condition of suddenness satisfies. In a similar phenomenon, if the nuclei receives a sudden jolt in its normal state, it may lead to the ionization or excitation of electrons in the atomic system. During fast ion-atom collisions target or projectile nuclei receives similar type of sudden jolt which can result in sudden change of the position of respective nuclei and thus creates a sudden perturbation in atomic electronic configuration to initiate the shaking processes. The sudden perturbation in electronic environment solely depends on the recoil of the target atom or incident projectile ion and therefore it affects also the projectile ion in same manner as target atom. The basic formulation to calculate shaking probability under the sudden jolt of nuclei is described in the book by Landau and Lifshitz \citep{Landau1980} using nonrelativistic hydrogenic wave functions. But the work is only limited to the hydrogen atom. The present work is an extension of the previous work to define the general expression for shaking probability for K-, L- and M- shell electrons of any atom using the hydrogenic wavefunctions due to the sudden jolt of nuclei during ion-atom collisions.

\section{Formalism of the Shaking Process}
\label{sec2}

Basically, shaking is a two-step process, in the first step due to the sudden perturbation, the central potential of parent atomic system or electronic environment gets disturbed, whereas in the second step under such influence electrons make transition to a new state (shakeup) or continuum state (shakeoff). In the case of ion-atom collisions, first step corresponds to sudden impact of projectile nuclei to target nuclei and second step resembles the shaking of electrons due to recoil of respective nuclei. Worth to note that, shaking due to sudden jolt of nuclei strongly depends on the first step or the amplitude of ion-atom impact, whereas shaking due to other processes e.g. inner shell ionization \citep{Mukoyama2005}, photoionization \citep{Mukoyamaa2010} etc. does not depend on the physical nature of first step. 

The shaking probability  of the system from one state to another due to sudden jolt can be determined from the general quantum mechanics rules, by the overlap integral of the corresponding wave functions \citep{Landau1980,Zettili,Bohm1979}
\begin{equation}\label{eq1}
W_{fi}=\left|\int \psi^{*}_{f}\psi_{i}^{(0)}dq\right|^{2}
\end{equation}
The both wave functions $\psi_{i}^{(0)}$ and $\psi_{f}$ are stationary and related to the Hamiltonians $\hat{H}_{0}$ and $\hat{H}$, respectively. Due to the stationary nature, the wavefunctions associated with the probability density is independent of time and have the form $\Psi$=$\psi$(q)$\exp (-iEt/\hbar)$, where the wave function depends only on the position coordinates and independent of the energy. As a result of the suddenness of perturbation, the electronic wave function of the system "has no time" to change from initial to final state and therefore remains intact as they were before the perturbation. Therefore they will no longer represent the characteristic wave functions of the new Hamiltonian of the system and will not corresponds the stationary states. In another words, shaking phenomena is the imperfect overlap of the initial and final electronic wave functions. For the case of non-relativistic hydrogenic wavefunctions Eq. \ref{eq1} can be rewritten as
\begin{equation}\label{eq2}
W_{fi}=\left|\int \psi^{*}_{n'l'm'}\psi_{nlm}dq\right|^{2}
\end{equation}
where n, l, m and n', l', m' are the quantum numbers of the initial and final atomic system, respectively. Important to note that, since the shake process has a monopole character, the selection rule for shaking processes only favours such transitions in which the principal quantum number of the final state is different from the initial i.e. n'$\neq$n and all other quantum numbers remain same i.e. l'=l and m'=m. Thus the Eq. \ref{eq2} can be simplified to

\begin{equation}\label{eq3}
W_{fi}=\left|\int \psi^{*}_{n'lm}\psi_{nlm}dq\right|^{2}
\end{equation}
\section{Theoretical work, Results and Discussion}
Total shaking probability of target atom or projectile ion due to the ion-atom collision can be calculated using the Eq. \ref{eq3}. Normalized wave functions corresponding to K-, L- and M- shells for an "infinitely heavy nucleus" \citep{Bransden2003} given in Table \ref{tab:table1} are used to calculate the integral of Eq. \ref{eq3}.
 \begin{table}[!h] 
\caption{\label{tab:table1} Normalized non-relativistic hydrogenic wave functions (in real forms) corresponding to the first three shells K-, L- and M- shells electrons. The spectroscopic notation s have their usual meanings.}
\vspace{2mm} 

\begin{longtable}{c l c l}

\hline
Shell  & Quantum & Spectroscopic & Wavefunction $\psi_{nlm}(r,\theta,\phi)$ \\
 & Numbers & Notation & \\
\hline
\hline
\\
 K & 1 0 0 & 1s & $\frac{1}{\sqrt{\pi }}\left(\frac{Z}{a}\right)^{3/2} \exp \left(-\frac{Zr}{a}\right)$\\\\
L & 2 0 0 &  2s  & $\frac{1}{2 \sqrt{2 \pi }}\left(\frac{Z}{a}\right)^{3/2} \left(1-\frac{Z r}{2 a}\right) \exp \left(-\frac{Z r}{2 a}\right)$ \\\\
 & 2 1 0 &  2p$_{z}$  & $\frac{1}{4 \sqrt{2 \pi } a}\left(\frac{Z}{a}\right)^{3/2}  \frac{Z r}{a} \exp \left(-\frac{Z r}{2 }\right)\cos \theta $\\
  & 2 1 1 &  2p$_{y}$  & $\frac{1}{8 \sqrt{\pi } } \left(\frac{Z}{a}\right)^{3/2}  \frac{Zr}{a} \exp \left(-\frac{Z r}{2 a}\right)\sin\theta \cos \phi$\\
   & 2 1 -1 &  2p$_{x}$  & $\frac{1}{8 \sqrt{\pi } } \left(\frac{Z}{a}\right)^{3/2}  \frac{Zr}{a} \exp \left(-\frac{Z r}{2 a}\right)\sin\theta \sin \phi$\\\\
M & 3 0 0 & 3s  &  $\frac{1}{3 \sqrt{3 \pi }}\left(\frac{Z}{a}\right)^{3/2} \left(\frac{2 r^2 Z^2}{27 a^2}-\frac{2 Z r}{3 a}+1\right) \exp \left(-\frac{Z r}{3 a}\right)$\\\\
 & 3 1 0  & 3p$_{z}$ & $\frac{2 \sqrt{2}}{27 \sqrt{\pi } } \left(\frac{Z}{a}\right)^{3/2}  \frac{Zr}{a} \left(1-\frac{Z r}{6 a}\right) \exp \left(-\frac{Z r}{3 a}\right)\cos \theta $\\

& 3 1 1 & 3p$_{y}$ & $\frac{2}{27 \sqrt{\pi } a} \left(\frac{Z}{a}\right)^{3/2}  \frac{Zr}{a} \left(1-\frac{Z r}{6 a}\right) \exp \left(-\frac{Z r}{3 a}\right)\sin\theta \cos \phi$\\
& 3 1 -1 & 3p$_{x}$ & $\frac{2}{27 \sqrt{\pi } a} \left(\frac{Z}{a}\right)^{3/2}  \frac{Zr}{a} \left(1-\frac{Z r}{6 a}\right) \exp \left(-\frac{Z r}{3 a}\right)\sin\theta \sin \phi$\\\\

& 3 2 0 & 3d$_{z^2}$& $\frac{1}{81 \sqrt{6 \pi } }\left(\frac{Z}{a}\right)^{3/2} \left(3 \cos ^2\theta -1\right) \left(\frac{Z^2 r^2 }{a^2}\right) \exp \left(-\frac{Z r}{3 a}\right)$\\

& 3 2 1 & 3d$_{yz}$ & $\frac{1}{81 \sqrt{\pi }}\left(\frac{Z}{a}\right)^{3/2}   \left(\frac{Z^2 r^2 }{a^2}\right) \exp \left(-\frac{Z r}{3 a}\right) \sin \theta  \cos \theta \cos \phi$\\
& 3 2 -1 & 3d$_{xz}$ & $\frac{1}{81 \sqrt{\pi }}\left(\frac{Z}{a}\right)^{3/2}  \left(\frac{Z^2 r^2 }{a^2}\right) \exp \left(-\frac{Z r}{3 a}\right) \sin \theta  \cos \theta  \sin \phi$\\
& 3 2 2 & 3d$_{x^2-y^2}$ & $\frac{1}{162 \sqrt{\pi }}  \left(\frac{Z}{a}\right)^{3/2}  \left(\frac{Z^2 r^2 }{a^2}\right) \exp \left(-\frac{Z r}{3 a}\right)\sin ^2\theta  \cos 2\phi$\\
& 3 2 -2 & 3d$_{xy}$ & $\frac{1}{162 \sqrt{\pi }}  \left(\frac{Z}{a}\right)^{3/2}  \left(\frac{Z^2 r^2 }{a^2}\right) \exp \left(-\frac{Z r}{3 a}\right)\sin ^2\theta  \sin 2\phi$\\\\

	\hline

\end{longtable}

\end{table}

It is worth to mention that after the sudden change occurring in the central potential or electronic environment, each electron has three possibilities in the new environment. It may remain either in the same state or can make a transition to unoccupied bound state (shakeup) or gets ionized to the continuum state (shakeoff). It is already mentioned in previous section that shaking of atom/ion is the sum of shakeup and shakeoff processes. So better way to calculate shaking probability is to subtract the probability that all the electron will remain in their initial state from the total probability i.e. unity. This method was applied by the earlier workers \citep{Mukoyama2005,Carlson1968} to calculate the shaking probabilities following the inner-shell vacancy production.

Now, during the ion-atom collisions the nucleus of the target atom/ projectile ion receives an impact which gives it a recoil velocity v. If this perturbation i.e. impact is sudden relative to electron orbital periods, it can lead to the excitation or ionization of the electrons \citep{Landau1980}.
Now, let us first discuss about sudden jolt in the target nuclei, assuming lab frame of reference is S. After the ion-atom collision the frame of reference is S', which is moving with the nucleus. Due to the suddenness of perturbation the coordinates of electrons in S' are same as S. The initial wave function of electron in S' is given by
\begin{equation}\label{eq4}
\psi_{0}^{'}=\psi_{0}\exp \left(-i q .\sum_{p} r_{p}  \right)
\end{equation}
Here, q = mv/$\hbar$ , v = atom/ion recoil velocity, m = atom/ion mass, summation is over all Z electrons in the atom and
$\psi_{0}$ is the wave function of electron when the nucleus is at rest or in frame S. 

From the Eq. \ref{eq3}, the required probability of electron to remain in the same state is given by
\begin{equation}\label{eq5}
W = \left|\int \psi_{0}^{2} \exp (-iq.r) dV\right| = \left|\int_0^\infty \int_0^\pi \int_0^{2\pi} \psi_{0}^{2} \exp (-iqr \cos \theta) r^{2} \sin \theta dr d\theta d\phi\right|
\end{equation}
Similar formalism is also applicable for projectile nuclei. Using Table \ref{tab:table1}, we can get the probability of electron to remain in same state (K-, L- and M- shell) after the sudden jolt of respective nuclei by integration of Eq. \ref{eq5}.
\\
The results obtained are following
\begin{equation}\label{eq6}
W^K_{1 0 0} = \frac{256 Z^8}{\left(a^2 q^2+4 Z^2\right)^4}
\end{equation}
\begin{equation}\label{eq7}
W^L_{2 0 0} = \frac{Z^8 \left(2 a^4 q^4-3 a^2 q^2 Z^2+Z^4\right)^2}{\left(a^2 q^2+Z^2\right)^8}
\end{equation}
\begin{equation}\label{eq8}
W^L_{2 1 0} = \frac{Z^{12} \left(Z^2-5 a^2 q^2\right)^2}{\left(a^2 q^2+Z^2\right)^8}
\end{equation}
\begin{equation}\label{eq9}
W^L_{2 1 1} = W^L_{2 1 -1} = \frac{4 Z^{12}}{\left(a^2 q^2+Z^2\right)^6}
\end{equation}

\begin{equation}\label{eq11}
W^M_{3 0 0} = \frac{256 Z^8 \left(3 a^2 q^2-4 Z^2\right)^2 \left(27 a^2 q^2-4 Z^2\right)^2 \left(243 a^4 q^4-216 a^2 q^2 Z^2+16 Z^4\right)^2}{\left(9 a^2 q^2+4 Z^2\right)^{12}}
\end{equation}
\begin{equation}\label{eq12}
W^M_{3 1 0} = \frac{65536 Z^{12} \left(3645 a^6 q^6-3321 a^4 q^4 Z^2+648 a^2 q^2 Z^4-16 Z^6\right)^2}{\left(9 a^2 q^2+4 Z^2\right)^{12}}
\end{equation}
\begin{equation}\label{eq13}
W^M_{3 1 1} = W^M_{3 1 -1} = \frac{262144 Z^{10} \left(81 a^4 q^4 Z-27 a^2 q^2 Z^3+4 Z^5\right)^2}{\left(9 a^2 q^2+4 Z^2\right)^{10}}
\end{equation}

\begin{equation}\label{eq15}
W^M_{3 2 0} = \frac{65536 Z^{14} \left(1377 a^4 q^4 Z-312 a^2 q^2 Z^3+16 Z^5\right)^2}{ \left(9 a^2 q^2+4 Z^2\right)^{12}}
\end{equation}
\begin{equation}\label{eq16}
W^M_{3 2 1} = W^M_{3 2 -1} = \frac{262144 Z^{14} \left(63 a^2 q^2 Z-4  Z^3\right)^2}{ \left(9 a^2 q^2+4 Z^2\right)^{10}}
\end{equation}

\begin{equation}\label{eq18}
W^M_{3 2 2} = W^M_{3 2 -2} = \frac{262144 Z^{16}}{ \left(9 a^2 q^2 + 4 Z^2\right)^8}
\end{equation}\newline 

\noindent where $W^K_{nlm}$, $W^L_{nlm}$ and $W^M_{nlm}$ are the probability that the electron will remain in same shell i.e. K-, L- and M- shell, respectively and a = Bohr radius and Z =  atomic number of the projectile or target according to the case.
For the case of Hydrogen atom Eq. \ref{eq6} reduces to the earlier reported value \citep{Landau1980} i.e. $W^K_{100} = 1/\left(1+\frac{a^2 q^2}{4}\right)^4$.
\\Thus the shaking probability is defined by
\begin{equation}
W_{shaking} = 1 - W
\end{equation}
where W can be replaced by $W^K_{1 0 0}$, $\sum_{2lm} W^L_{n l m}$ or $\sum_{3lm}W^M_{n l m}$ from Eq. \ref{eq6} - \ref{eq18} to get the shaking probability, $W_{shaking}$, for K-, L- and M- shell electrons respectively.
\\
It is found that for the values of a.q $\ll$ 1, shaking probability reduces to 0 whereas for the 
a.q $\gg$ 1 it equals to unity. Interestingly during the fast ion-atom collisions, condition a.q $\gg$ 1  is applicable in wide range of recoil energy (starts from keV). Noteworthy that applicability of the condition of suddenness during perturbation requires higher  incident energy in the ion-atom collisions than required by the recoil energy of respective nuclei for the shaking. It clearly suggests that once the sudden approximation condition is satisfied it will naturally assured the limiting condition a.q $\gg$ 1.

\section{Conclusion}
We have calculated the shaking probabilities accompanying the ion-atom collisions using the hydrogenic wavefunctions for K-, L- and M- shell electrons. It is found that in the limiting case i.e. a.q $\ll$ 1, the shaking probability tends to zero, whereas for a.q $\gg$ 1 it tends to unity. Interestingly, the condition a.q $\gg$ 1 holds good even for the low recoil cases (starting from keV/u). During the fast ion-atom collision suddenness of impact can satisfy this condition very well. Thus this study implies that during fast heavy ion-atom collisions, one of the most probable channel of electron transitions is the shaking (shakeup + shakeoff), which occurs due to sudden jolt of projectile/target nuclei in the collisions.

\section{Acknowledgement}
Prashant Sharma is thankful to UGC, India for providing the fellowship as financial support to carry out this work.

\section*{References}

\bibliography{mybibfile}

\begin{thebibliography}{}
\expandafter\ifx\csname url\endcsname\relax
  \def\url#1{\texttt{#1}}\fi
\expandafter\ifx\csname urlprefix\endcsname\relax\def\urlprefix{URL }\fi
\expandafter\ifx\csname href\endcsname\relax
  \def\href#1#2{#2} \def\path#1{#1}\fi

\end{thebibliography}


\begin{thebibliography}{10}
\expandafter\ifx\csname url\endcsname\relax
  \def\url#1{\texttt{#1}}\fi
\expandafter\ifx\csname urlprefix\endcsname\relax\def\urlprefix{URL }\fi
\expandafter\ifx\csname href\endcsname\relax
  \def\href#1#2{#2} \def\path#1{#1}\fi

\bibitem{Freedman1974}
M.~S. Freedman,
  \href{http://www.annualreviews.org/doi/abs/10.1146/annurev.ns.24.120174.001233}{{Atomic
  Structure Effects in Nuclear Events}}, Annual Review of Nuclear Science
  24~(1) (1974) 209--248.
\newblock \href {http://dx.doi.org/10.1146/annurev.ns.24.120174.001233}
  {\path{doi:10.1146/annurev.ns.24.120174.001233}}.
\newline\urlprefix\url{http://www.annualreviews.org/doi/abs/10.1146/annurev.ns.24.120174.001233}

\bibitem{ff}
K.~Hecht, \href{http://dx.doi.org/10.1007/978-1-4612-1272-0_59}{Sudden and
  adiabatic approximations}, in: Quantum Mechanics, Graduate Texts in
  Contemporary Physics, Springer New York, 2000, pp. 561--571.
\newblock \href {http://dx.doi.org/10.1007/978-1-4612-1272-0_59}
  {\path{doi:10.1007/978-1-4612-1272-0_59}}.
\newline\urlprefix\url{http://dx.doi.org/10.1007/978-1-4612-1272-0_59}

\bibitem{Feoktistov2010}
O.~I. Feoktistov, {Atomic ionization as a sudden perturbation of an electron by
  the charge of a projectile}, Ukrainian Journal of Physics 55 (2010) 165--169.

\bibitem{Levinger1953}
J.~S. Levinger, \href{http://link.aps.org/doi/10.1103/PhysRev.90.11}{Effects of
  radioactive disintegrations on inner electrons of the atom}, Phys. Rev. 90
  (1953) 11--25.
\newblock \href {http://dx.doi.org/10.1103/PhysRev.90.11}
  {\path{doi:10.1103/PhysRev.90.11}}.
\newline\urlprefix\url{http://link.aps.org/doi/10.1103/PhysRev.90.11}

\bibitem{Couratin2012}
C.~Couratin, P.~Velten, X.~Fl\'{e}chard, E.~Li\'{e}nard, G.~Ban, A.~Cassimi,
  P.~Delahaye, D.~Durand, D.~Hennecart, F.~Mauger, A.~M\'{e}ry,
  O.~Naviliat-Cuncic, Z.~Patyk, D.~Rodr\'{\i}guez, K.~Siegień-Iwaniuk, J.-C.
  Thomas, \href{http://link.aps.org/doi/10.1103/PhysRevLett.108.243201}{{First
  Measurement of Pure Electron Shakeoff in the $\beta$ Decay of Trapped
  \^{}\{6\}He\^{}\{+\} Ions}}, Physical Review Letters 108~(24) (2012) 243201.
\newblock \href {http://dx.doi.org/10.1103/PhysRevLett.108.243201}
  {\path{doi:10.1103/PhysRevLett.108.243201}}.
\newline\urlprefix\url{http://link.aps.org/doi/10.1103/PhysRevLett.108.243201}

\bibitem{Sharma2015}
P.~Sharma, T.~Nandi, {Theoretical Studies on Shakeoff Processes in Nuclear
  Reactions} (2015) (To be Submitted).

\bibitem{Cho1997}
H.~J. Cho, S.~K. Nha, {Shake-off Probability for Electron Capture Decay of
  207Bi}, Journal of the Korean Physical Society 30~(3) (1997) 521--527.

\bibitem{PhysRevC.11.1353}
T.~Mukoyama, S.~Shimizu,
  \href{http://link.aps.org/doi/10.1103/PhysRevC.11.1353}{{Electron shakeoff
  accompanying internal conversion}}, Phys. Rev. C 11~(4) (1975) 1353--1363.
\newblock \href {http://dx.doi.org/10.1103/PhysRevC.11.1353}
  {\path{doi:10.1103/PhysRevC.11.1353}}.
\newline\urlprefix\url{http://link.aps.org/doi/10.1103/PhysRevC.11.1353}

\bibitem{PhysRevC.3.2246}
F.~T. Porter, M.~S. Freedman, F.~Wagner,
  \href{http://link.aps.org/doi/10.1103/PhysRevC.3.2246}{{Ionization (Shakeoff)
  Accompanying K-Shell Internal Conversion}}, Phys. Rev. C 3~(6) (1971)
  2246--2259.
\newblock \href {http://dx.doi.org/10.1103/PhysRevC.3.2246}
  {\path{doi:10.1103/PhysRevC.3.2246}}.
\newline\urlprefix\url{http://link.aps.org/doi/10.1103/PhysRevC.3.2246}

\bibitem{Mukoyamaa2010}
T.~Mukoyamaa, {Saturation energy of shake process in photoionization}, X-Ray
  Spectrometry 39~(November 2009) (2010) 142--145.
\newblock \href {http://dx.doi.org/10.1002/xrs.1232}
  {\path{doi:10.1002/xrs.1232}}.

\bibitem{Mukoyama2005}
T.~Mukoyama, \href{http://doi.wiley.com/10.1002/xrs.762}{{Electron shake
  process following inner-shell ionization}}, X-Ray Spectrometry 34~(1) (2005)
  64--68.
\newblock \href {http://dx.doi.org/10.1002/xrs.762}
  {\path{doi:10.1002/xrs.762}}.
\newline\urlprefix\url{http://doi.wiley.com/10.1002/xrs.762}

\bibitem{Cho19972}
H.-j. Cho, S.-k. Nha, G.-d. Kim, {Study of the nuclear shakeoff in (EC+
  $\beta$+) decay nucleus}, Journal of the Korean Physical Society 30~(2)
  (1997) 180--185.

\bibitem{PhysRevC.25.2722}
A.~Suzuki, J.~Law,
  \href{http://link.aps.org/doi/10.1103/PhysRevC.25.2722}{Electron capture
  shakeoff}, Phys. Rev. C 25 (1982) 2722--2727.
\newblock \href {http://dx.doi.org/10.1103/PhysRevC.25.2722}
  {\path{doi:10.1103/PhysRevC.25.2722}}.
\newline\urlprefix\url{http://link.aps.org/doi/10.1103/PhysRevC.25.2722}

\bibitem{Landau1980}
L.~Landau, E.~Lifshitz, {Quantum Mechanics - Non-relativistic Theory}, Vol.~1,
  Pergamon Press, New York, 1980.

\bibitem{Zettili}
N.~Zettili, {Quantum Mechanics Concepts and Applications}, 2nd Edition, John
  Wiley \& Sons, Ltd., 2009.

\bibitem{Bohm1979}
A.~Bohm, {Quantum mechanics.}, Springer, New York, 1979.

\bibitem{Bransden2003}
B.~H. Bransden, C.~J. Joachain, {Physics of atoms and molecules}, Longman,
  London; New York, 1983.

\bibitem{Carlson1968}
T.~A. Carlson,
  \href{http://link.aip.org/link/?JCP/48/5191/1\&Agg=doi}{{Measurement of the
  Relative Abundances and Recoil Energy Spectra of Fragment Ions Produced as
  the Initial Consequences of X-Ray Interaction with C2H5I, CH3CD2I, and
  Pb(CH3)4}}, The Journal of Chemical Physics 48~(11) (1968) 5191.
\newblock \href {http://dx.doi.org/10.1063/1.1668193}
  {\path{doi:10.1063/1.1668193}}.
\newline\urlprefix\url{http://link.aip.org/link/?JCP/48/5191/1\&Agg=doi}

\end{thebibliography}

\end{document}